\documentclass[aps,prl,twocolumn,amsmath, amssymb,graphicx,superscriptaddress]{revtex4-1}

\usepackage{bm}
\usepackage{subfigure}
\usepackage[dvips,xetex]{graphicx}
\usepackage{amsthm}
\usepackage{notes2bib}
\usepackage{amsmath}
\usepackage{amssymb}
\usepackage{url}
\usepackage{enumerate}
\usepackage{epstopdf}
\usepackage{color}
\usepackage{mathtools}          
\usepackage{setspace}

\usepackage[usenames,dvipsnames,svgnames,table]{xcolor}
\usepackage{color}

\newcommand{\R}{{\mathcal{R}}}

\begin{document}

\title{The continuous route to multi-chaos}

\author{Yoshitaka Saiki}
\affiliation{Graduate School of Commerce and Management, Hitotsubashi University, Tokyo 186-8601, Japan}
\affiliation{JST, PRESTO, Saitama 332-0012, Japan}
\affiliation{University of Maryland, College Park, Maryland 20742, USA}
\author{Miguel A.F. Sanju\'{a}n}
\affiliation{University of Maryland, College Park, Maryland 20742, USA}
\affiliation{Nonlinear Dynamics, Chaos and Complex Systems Group, Departamento de F\'{i}sica\\ Universidad Rey Juan Carlos, Tulip\'{a}n s/n, 28933 M\'{o}stoles (Madrid) Spain}
\author{James A. Yorke}
\affiliation{University of Maryland, College Park, Maryland 20742, USA}

\begin{abstract}
For low-dimensional chaotic attractors
there is usually a single number of unstable dimensions for all of its periodic orbits and we can say such attractors exhibit ``mono-chaos''. 
In high-dimensional chaotic attractors, trajectories are prone to travel through quite different regions of phase space, some far more unstable than others. This heterogeneity makes predictability even more difficult than in low-dimensional homogeneous chaotic attractors. 
A chaotic attractor is ``multi-chaotic'' if every point of the attractor is arbitrarily close to periodic points with different numbers of unstable dimensions. 
We believe that most physical systems possessing a high-dimensional attractor are of this type.
We make three conjectures about multi-chaos which we explore using three two-dimensional paradigmatic examples of multi-chaotic attractors. 
They can be thought of as small-scale examples that give insight for real high-dimensional phenomena. 
We find a single route from mono-chaos to multi-chaos if an attractor changes continuously as a parameter is varied.
This multi-chaos bifurcation (MCB) is a periodic orbit bifurcation; one branch of periodic orbits is created with a  number of unstable dimensions that is different from the mono-chaos. 
\end{abstract}

\date{\today}
\maketitle

{\bf Introduction.} 
Prediction for chaotic systems occurs throughout science. For example forecasting geomagnetic storms and solar flares \cite{pariat_2017} or natural hazards \cite{guzzetti_2016} or earthquakes \cite{tian_2017} or weather \cite{patil_2001}. 
Predictability is more difficult when the ``chaotic attractor''
\bibnote{A set $S$ is a {\bf chaotic attractor} if (1) it is {\bf invariant} (if a trajectory is in $S$ at some time, then it is in $S$ for all later time), (2) $S$ has a dense trajectory with at least one positive Lyapunov exponent, and (3) trajectories near $S$ are attracted to it as time increases.}, 
is heterogeneous, i.e. if different regions of the chaotic attractor are unstable in more directions than in others.
Predictability is especially difficult when a trajectory enters a region that has more unstable directions than the region it is leaving. Then ``shadowing'' breaks down: numerical simulations no longer reflect true behavior--as we explain later.

In our work with simple whole earth weather models  (e.g., \cite{patil_2001}), the phase space had dimension $3$$\times$$10^6$, trajectories were chaotic, and we estimate that there were $3$$\times$$10^4$ unstable directions, that is a tiny ellipse around an initial point would expand in $3$$\times$$10^4$ dimensions. For regional short-term prediction, the number of unstable directions is effectively reduced by perhaps a factor of 100, but for storm conditions it is higher and thus prediction is much more difficult. 

If the approximate state of the weather is known, (some point $x$ in phase space), and the unstable dimension of the dynamics is $D$ at that point, then
after a few hours the possible weather states lie on an expanding ellipse of dimension $D$. 
To update the current state of the weather every few hours, it suffices to determine the location on that ellipsoid. The number of data points (point measurements of temperature, humidity, pressure, etc) needed for that is proportional to $D$.

\begin{figure}
  \includegraphics[width=.45\textwidth]{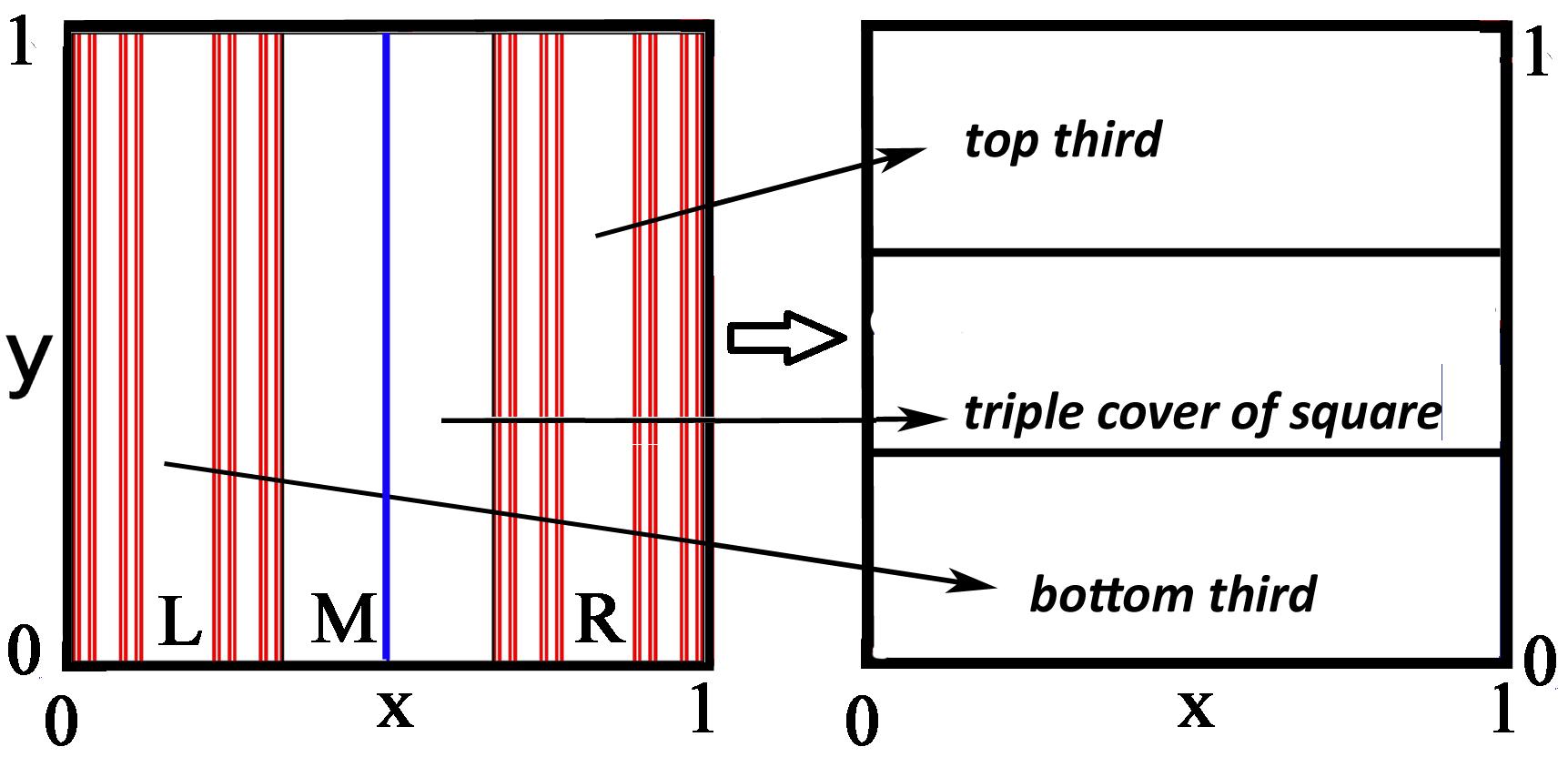}
  \caption{
    {\bf Our Multi-Chaos Baker Map $B_{MC}(x,y)$}. 
The figure shows a three-piece version of the baker map. 
We divide $0\le x<1$ into three intervals,
 $L=[0,1/3),M=[1/3,2/3),$ and $R=[2/3,1)$,  and divide the square into $3$ tall rectangles $\R_L, \R_M,$ and $\R_R$
 whose bases are $L, M,$ and $R$. 
The map  $B_{MC}$ is defined as follows:
For $x\in L,~y\mapsto y/3.$ 
For $x\in M,~y\mapsto 3y \bmod 1.$ 
For $x \in R,~y\mapsto y/3 + 2/3.$ 
And then $x\mapsto 3x \bmod1$. 
Hence $B_{MC}$ expands each rectangle horizontally to full width as shown. The region $\R_1$ = $\R_L \cup\R_R$ is contracted vertically. The region $\R_2=\R_M$ is expanded in both coordinates  so that its image is a triple cover of the entire square. Hence $\R_1$ and $\R_2$ are regions of one- and two-dimensional instability.
The vertical red lines constitute the invariant set whose trajectories stay in  $\R_1$ and the vertical blue line (at $x=1/2$) is the invariant set of points whose trajectories stay inside  $\R_2$. }
\label{fig:baker}
\end{figure}

Simple models having regions of phase space with different numbers of unstable dimensions have been lacking in the literature, and here we introduce three simple examples as prototypes for understanding far higher dimensional situations. Our first example is shown in Fig.~\ref{fig:baker}.
There are two regions ($\R_1 =$ the left and right thirds of the square) where the dynamics is unstable in one direction (the $x$ coordinate) while in the middle third, denoted $\R_2$, it is unstable in both $x$ and $y$ coordinates.
There exist trajectories that stay in either region but almost every trajectory wanders through the entire square.
 $\R_1$ has a fixed point at $(0,0)$ whose unstable dimension is 1 and $\R_2$ has a fixed point $(1/2,1/2)$ that has unstable dimension 2.
When an attractor has 2 periodic orbits that are unstable in different numbers of dimensions, we say the attractor has {\bf Unstable Dimension Variability (UDV)} \cite{kostelich_1997}.
 In contrast if all trajectories in a chaotic attractor are unstable in the same number of directions, we say there is {\bf mono-chaos}. 
 
In this paper we present evidence that UDV attractors are quite strange, that they are ``multi-chaotic''. If a periodic orbit is unstable in $k$ directions, 
we say {\bf it has UD-$k$}.

{\bf Conjecture~1.} If there is one UD-$k$ periodic orbit, then 
almost always there are infinitely many UD-$k$ periodic points and they lie arbitrarily close to every point of the attractor. 

A chaotic attractor has {\bf multi-chaos} if arbitrarily close to each point of the attractor there are $D$-dimensionally unstable periodic points and this is true for multiple values of $D$.
We expect that most high-dimensional attractors are multi-chaotic. 
It can be shown that the chaotic attractor is in fact multi-chaotic for each of our examples. For our Multi-Chaos Baker map, Fig.~\ref{fig:baker},  the chaotic attractor is the entire square and 
saddle  periodic points are dense in the square as are repelling periodic points.
 
 In addition to presenting low-dimensional examples, the purpose of this paper is ask how UDV arises from mono-chaos as some physical parameter is varied. We believe if an attractor is changing continuously, the transition will occur at a periodic orbit bifurcation and we give some examples of this transition.

{\bf Multi-chaos.}
In our two-dimensional examples, UD-$1$ orbits are saddles and UD-$2$ orbits are repellers. Hence if an attractor (with a dense trajectory) has a UD-$1$ orbit and a UD-$2$ orbit, the attractor has UDV.

   A consequence of UDV is that any trajectory that wanders densely through the invariant set will occasionally get very close to each point of each periodic orbit. Therefore that trajectory will spend arbitrarily long intervals of time near each of the fixed points (or periodic orbits).  
Hence for each time $T>0$ the trajectory's time-$T$ positive Lyapunov exponents
will occasionally be  the same as for the periodic orbit it approaches. 

A UDV attractor has {\bf  multi-chaos}  \cite{das_2017a} if it has two (or more) dense sets of periodic orbits with different UD values.

{\bf Conjecture~2.} UDV always implies multi-chaos.

{\bf Remark.}
Multi-chaos should not be confused with ``hyper-chaos''.
A multi-chaotic attractor can have one or more positive Lyapunov exponents. It need not be hyper-chaotic  (i.e., having more than one positive Lyapunov exponent). 
Furthermore all periodic orbits of a hyper-chaotic attractor might  have the same UD value, in which case it would not be multi-chaotic.

{\bf The crisis route to  multi-chaos.}
As some parameter, say $\alpha$, is varied, a ``crisis'' is a sudden discontinuous change in the size of a chaotic attractor, at some value $\alpha_0$. 
Hence, a crisis can be seen as a sudden jump in the plot of an attractor versus $\alpha$.
On the side of $\alpha_0$ where the attractor is small, the attractor could be mono-chaotic. On the other side, the attractor has included a large region of phase space which may include a periodic orbit of a different UD value. Then the attractor is multi-chaotic. See \cite{alligood_2006, das_2017a, viana_2005, pereira_2007}.

{\bf  The continuous route to  multi-chaos.} 
If as a parameter $\alpha$ is varied,
a mono-chaotic attractor suddenly becomes multi-chaotic after some $\alpha =\alpha_{MC}$, we say a   {\bf multi-chaos bifurcation (MCB)} occurs at $\alpha_{MC}$. 
What is the nature of this bifurcation?
As a parameter changes, a periodic orbit in a chaotic attractor can migrate to a region that is more unstable, and the orbit's UD value can increase. Then an exponent of that orbit will pass through $0$ and a bifurcation will occur. 
Or a new pair of orbits can appear in an analogue of a saddle-repeller bifurcation, with UD values $k$ and $k+1$ for some $k>0$.

{\bf Conjecture~3.}  
 For a typical attractor, if an MCB 
occurs as the attractor changes continuously (without a 
crisis), then there 
will be 
a periodic orbit bifurcation, i.e.,
either period-doubling or pitchfork or Hopf or  pair-creation such as saddle-repeller. 

{\bf Expanding regions $\R_k$ and ``index sets''.}
Let $\R_k$ denote the region of phase space in which the dynamics (specifically, the map's Jacobian) is $k$-dimensionally expanding; see e.g. Fig.~\ref{fig:baker}.
We call the largest invariant set that lies wholly in $\R_k$ the  {\bf index-$k$ set}.
For Fig.~\ref{fig:baker},
$\R_1$ and $\R_2$ are defined in the caption. At the center of 
Fig.~\ref{fig:two_cantor_sets}-Right, there is a different $\R_2$, the white rectangle ($1/3<x<2/3, -c <y<1+c$) where $c=(1-\alpha)/(\sigma-\alpha)\approx 0.13$, and $\R_1$ is the rest, excluding boundary points. 

 It probably seems strange that the existence of two periodic orbits with different UD values has such a dramatic consequence for an attractor that it implies multi-chaos.
Our response is that these orbits generally lie in index sets, that can be quite big as Figs. 
\ref{fig:baker} and 
\ref{fig:two_cantor_sets} illustrate.

{\bf Multi-chaos connects many phenomena like fluctuating exponents (FE) and  UDV.} 
Multi-chaotic attractors contain periodic orbits with different UDs. A typical trajectory will  return near each, occasionally spending long times near them before moving on, and while near the periodic orbit of a region, it will have the same number of positive finite-time Lyapunov exponents (FTLEs) as the periodic orbit. As it moves among the periodic orbits, 
its number of positive FTLEs fluctuates (for each time $T>0$);  see \cite{dawson_1994, dawson_1996}. This property is referred to as FE (Fluctuating Exponents). 
Some papers have used the term UDV to mean FE.
UDV and FE are both implied by other dynamical phenomena in the literature such as riddled basins,
blowout bifurcations, on-off intermittency, and chaotic itinerancy \cite{ott_1993, platt_1993, heagy_1994, ott_1994, tsuda_2009}.

Transitions from mono-chaos to FE or UDV have been observed in \cite{dawson_1996, moresco_1997, barreto_2000b}, but the mechanism of the transitions is not discussed.

{\bf Shadowing.} It is important for a physicist to know how good a numerical simulation is -- as in a climate simulation -- and for how long it is valid. 
When each numerical trajectory stays close to some actual trajectory of the system, we say the system has the {\bf shadowing property,} i.e. simulations are realistic. 

When a trajectory moves from a region where the dynamics has fewer unstable directions to a region where it has more, shadowing fails, and trajectories become unrealistic
-- see Fig.~3 of 
 \cite{grebogi_2002}. Such a transition causes fluctuations in the number of positive FTLEs,
 which means FE will be common in  higher-dimensional attractors.

The FE property implies shadowing fails, as was established by   \textcite{dawson_1994}. 
Mono-chaotic systems can have the shadowing property but multi-chaotic systems cannot, as shown for UDV in  \cite{sauer_1997,
yuan_2000, grebogi_2002}.

{\bf UDV in the mathematics literature.}
The first examples of a (robust) invariant set 
containing periodic orbits with different UD were given by \textcite{abraham_1970, simon_1972} in four and three dimensions, respectively.  ``Robust'' means the property persists under all sufficiently small perturbations. 
 Later it was mathematically studied using the notions of ``blenders'' and ``hetero-dimensional cycles'' (see \cite{bonatti_2005} and references therein).
That literature generally shows no interest in whether their invariant sets are (physically observable) attractors.

\begin{figure}[t]%
\includegraphics[width=.23\textwidth]{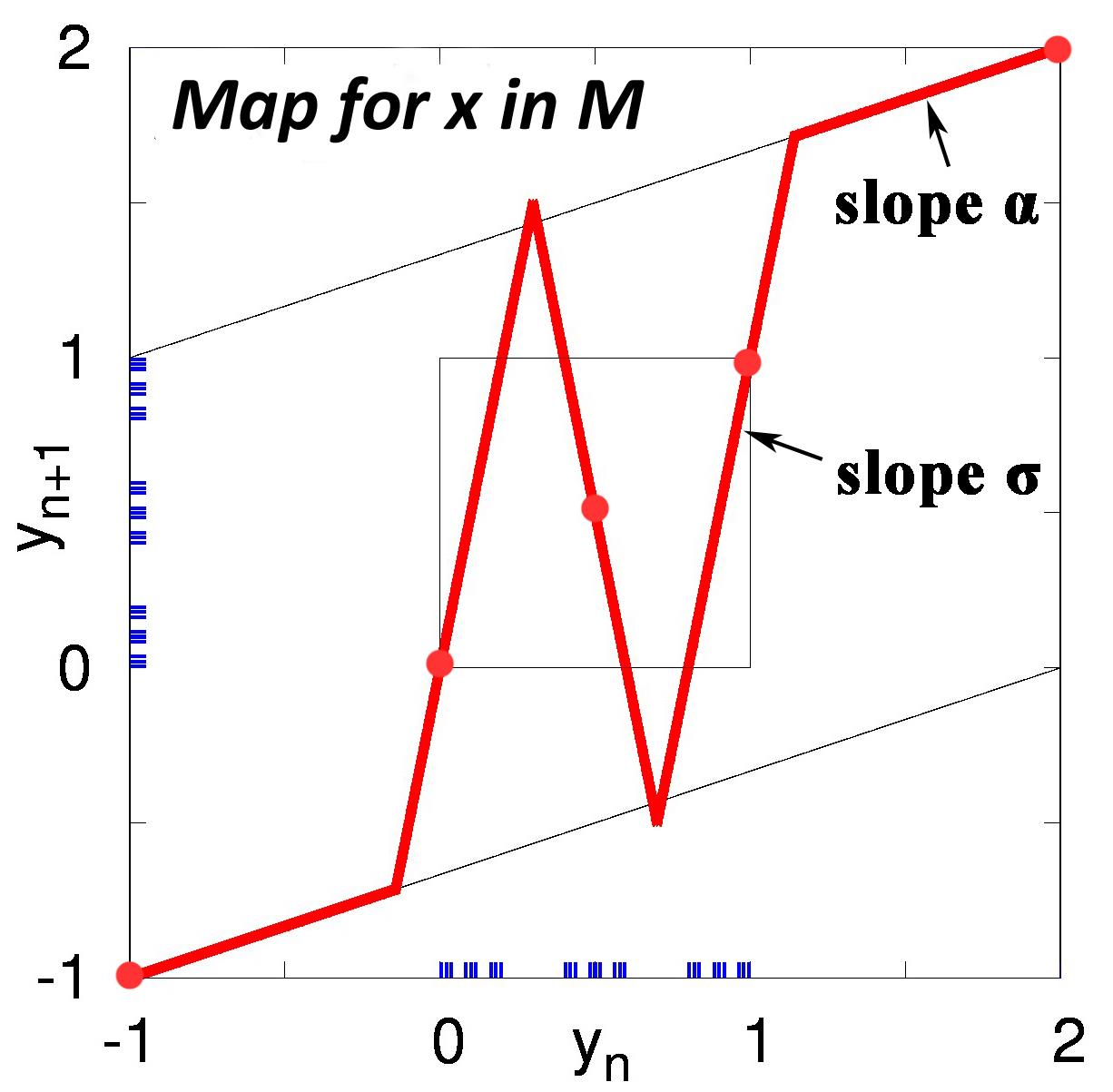}
\caption{{\bf Defining our Zigzag Map}. 
Here, as in Fig.~\ref{fig:baker}, 
the definition of the map depends on which of the three intervals $x$ is in: $L=[0,1/3),M=[1/3,2/3),$ and $R=[2/3,1)$. 
For $x\in L$,  
$y\mapsto -1+\alpha (y+1)$. 
For $x\in R$, 
$y\mapsto 2+\alpha (y-2)$. 
For $x\in M$, the 
figure shows the map. 
Each of the three maps is from [-1,2] into itself. 
The horizontal coordinate $x\mapsto 3x \bmod1$.
Each slope in the map shown is either $0<\alpha<1$ or $\pm\sigma$, where $\sigma>3$.
Here $\alpha = 1/3,$ and $\sigma=5$ both here and in Fig.~\ref{fig:two_cantor_sets}.
All 5 fixed points are shown with large red dots.
The Zigzag Map also has an invariant fractal set for $x=1/2$ ($x$ is not shown)  illustrated with dots on axes.
}
\label{fig:multi-chaos-baker}
\end{figure}
\begin{figure}	
  \includegraphics[width=.238\textwidth,height=.238\textwidth]{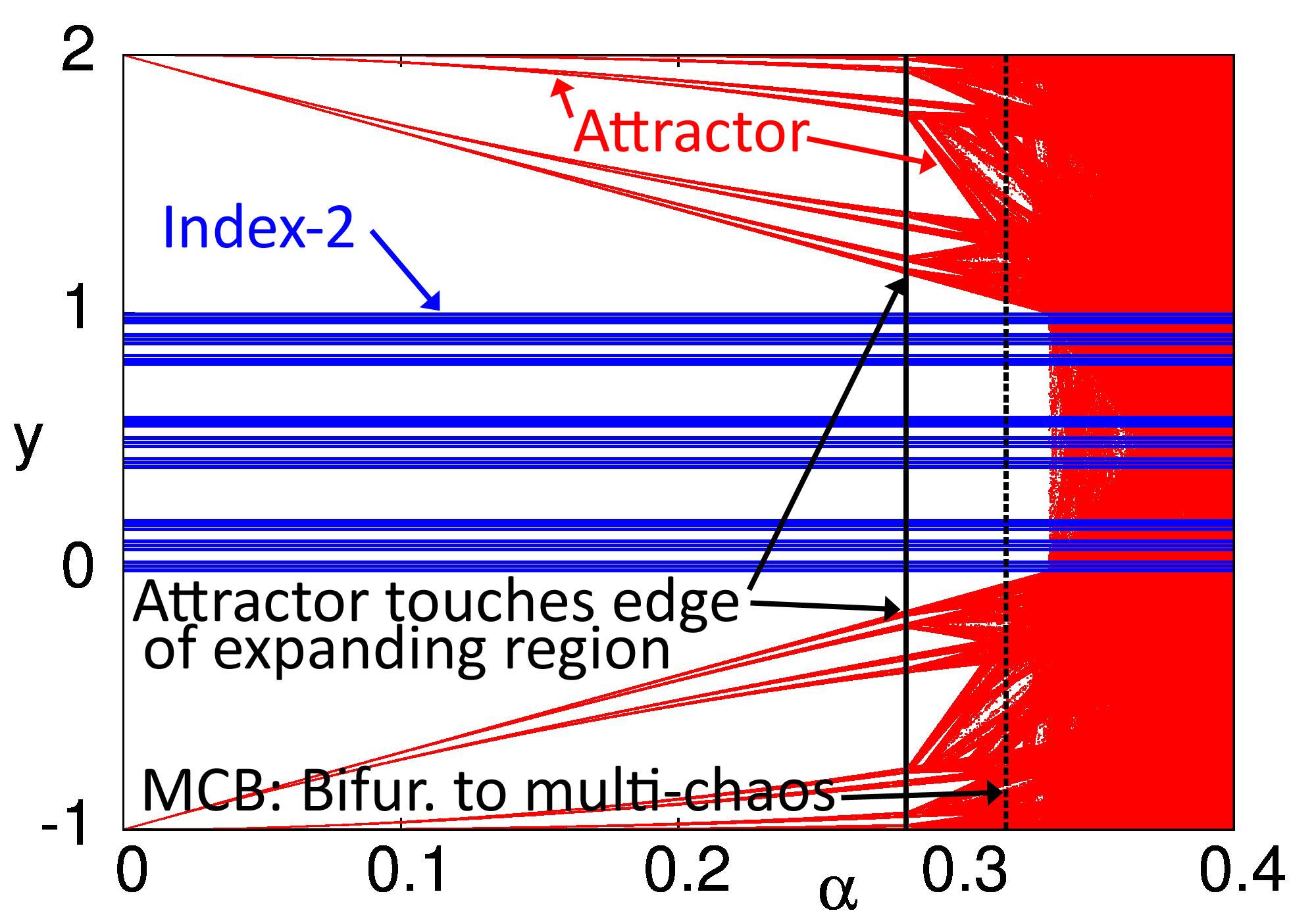}
  \includegraphics[width=.238\textwidth,height=.238\textwidth]{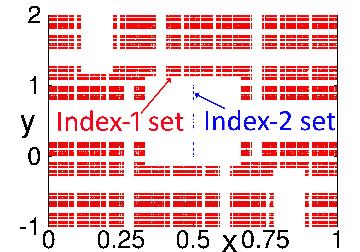}
  \caption{\textbf{The Zigzag Map's bifurcation diagram and index sets.}
{\bf Left panel.}  The chaotic attractor (red) is shown increasing in size as $\alpha$ increases.
The blue set is the index-$2$ set.     
At $\alpha\approx 0.28$ (solid black vertical line) the attractor begins to move into the expanding region, but the attractor does not contain repelling periodic points until after $\alpha_{MC}\approx 0.31$ (dotted black vertical line), when a period-4 repeller exists. Then there is multi-chaos. 
    At $\alpha=1/3$ there is a ``crisis'' after which the attractor jumps in size and is the entire $x$-$y$ square.
{\bf Right panel.}   Here $\alpha = 0.4~ (> 1/3)$. We show only 
the index-$1$ set (red) and the index-$2$ set (blue), which is on the vertical line $x=1/2$. This illustrates that within the multi-chaotic attractor (the entire square) there are relatively large index sets.
}\label{fig:two_cantor_sets}
\end{figure}
%


{\bf Our Multi-Chaos Maps.}
Our three maps (including the Multi-Chaos Baker map introduced above) have the following property. Each periodic orbit lying wholly in some $\R_k$ has UD-$k$. This is true because 
for each map $F$, the Jacobian matrix $DF(x,y)$ has the lower triangular form 
$\bigl(\begin{smallmatrix}3 & 0\\ c & d\end{smallmatrix}\bigr)$. 
The Jacobian $DF^T$ of the time-$T$ map $F^T$ is also lower triangular since by the chain rule, $DF^T$ is the product of $T$ of these matrices $DF$. The number of expanding directions for a point is the number of diagonal elements of $DF$ that are $>1$.

{\bf Our ``Zigzag'' Map and its route to multi-chaos. }
As with the Multi-Chaos Baker Map,  
the next two-dimensional map has $x$ dynamics described by $x\to 3x ~\bmod 1$,
and its $y$ dynamics depends on whether $x$ is in $L$, $M$, or $R$. It has two slope parameters, $\alpha$ and $\sigma$. 
 Figure~\ref{fig:multi-chaos-baker} shows the $y$ dynamics on $M$ and the caption gives the map also on $L$ and $R$.
The map has an index-$2$ fractal invariant set on the vertical line at $x=1/2$ for every $\alpha$ and every $\sigma>1$; (we use $\sigma = 5$ and then its
dimension is ${\ln 3}/{\ln 5}\approx 0.683$).
The attractor is chaotic for all $\alpha>0$, and for $\alpha<0.28$  is an index-$1$ set.

As $\alpha$ increases from $0$, at $\alpha_{MC}\approx 0.31$ (see
the left panel of Fig.~\ref{fig:two_cantor_sets}), there is   a pitchfork bifurcation of a period-$4$ periodic orbit, one of whose branches consists of repellers. Numerically this appears to be the first occurrence in the attractor of a repelling periodic orbit.
This observation supports Conjecture~3.
Hence the MCB occurs at $\alpha_{MC}$.

At $\alpha=1/3$, the attractor collides with the index-$2$ set, after which the attractor suddenly jumps in size, covering the whole $x$-$y$ square. 
For $\alpha=0.4$, both index-$1$ and index-$2$ sets coexist 
(see right panel of Fig.~\ref{fig:two_cantor_sets}).
We have identified the index sets by using the Stagger-and-Step method \cite{sweet_2001c}.

\begin{figure}
\includegraphics[width=.238\textwidth,height=.238\textwidth]
{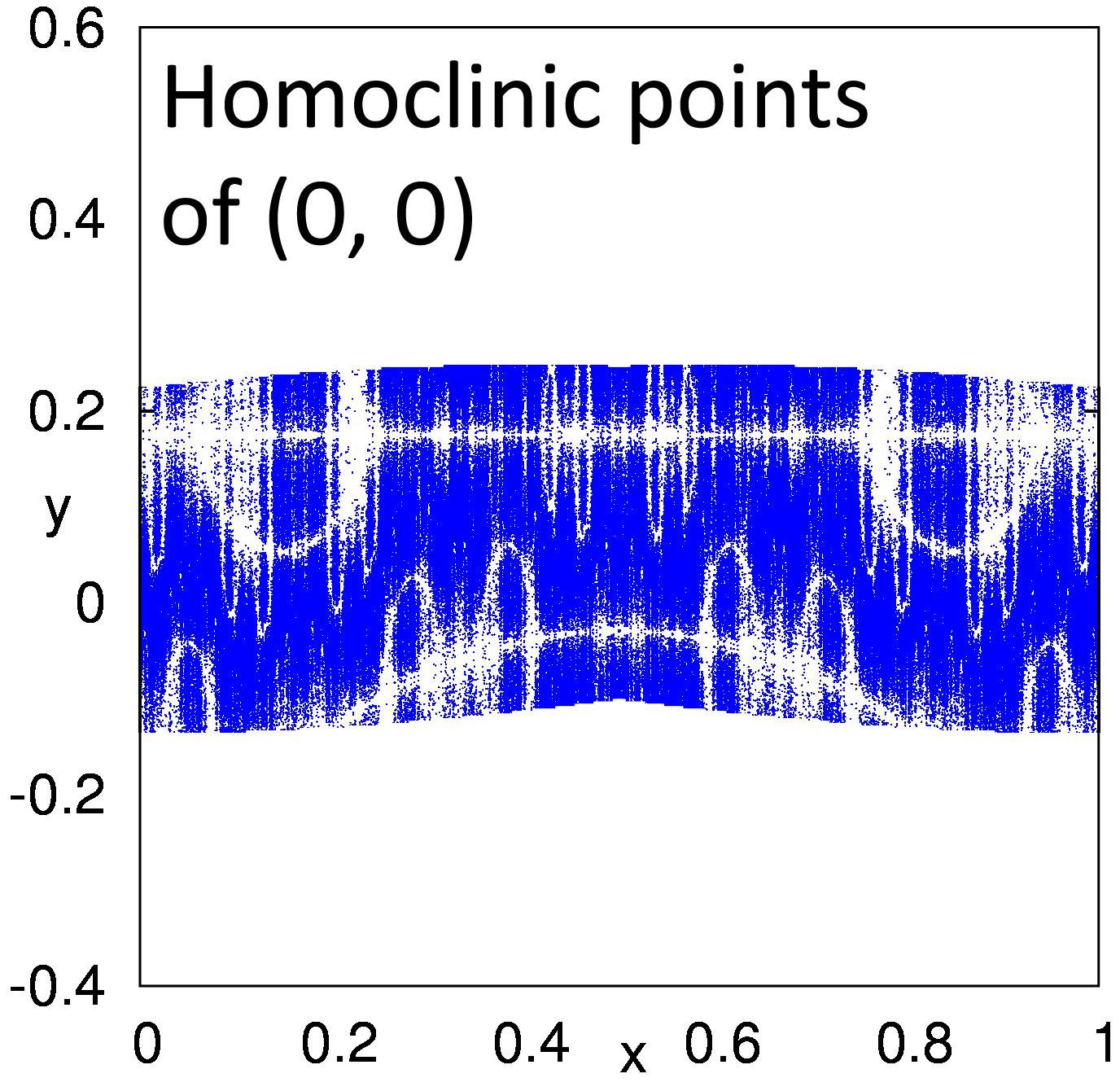}
\includegraphics[width=.238\textwidth,height=
.238\textwidth
]{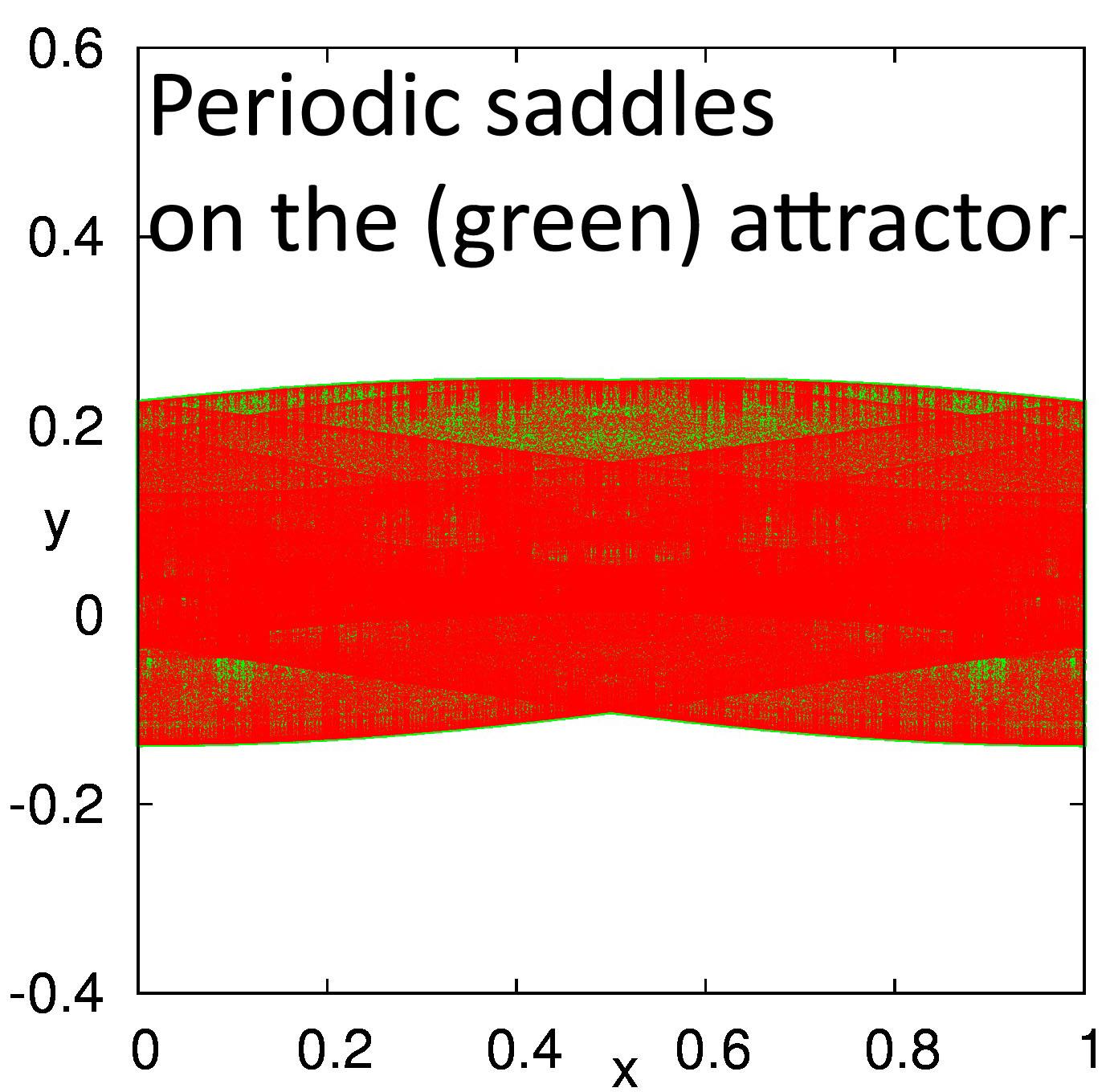}
\caption{ {\bf Homoclinic points and periodic saddles  for Kostelich map.}
It can be shown that both saddles and repellers are dense in the attractor, so that we have multi-chaos. This figure shows what the sets of homoclinic points and periodic points look like for limited computations. 
{\bf Left panel.} 
Points in the attractor that map to the repelling origin within $14$ iterates. Since they are in the unstable manifold of $(0,0)$, they are homoclinic points.
 In fact, the homoclinic points can be shown to be densely distributed in the attractor implying that repelling periodic points are dense, since  Marotto \cite{marotto_2005} shows that arbitrarily close to each homoclinic point there are repelling periodic points.
 {\bf Right panel.} 
 The saddle periodic points (red) of 
period $13$ 
are plotted on top of the chaotic attractor (green).
They become denser as the period increases. 
}
\label{fig:Kos-saddles}
\end{figure}

{\bf Kostelich map.} Consider a smooth map \cite{kostelich_1997,das_2017a}:
\begin{equation*}
\begin{array}{lcl}
x_{n+1}&=&3 x_n \bmod1\\
y_{n+1}&=&y_n -\sigma \sin(2 \pi y_n)+\alpha (1-\cos(2 \pi x_n))\bmod1. 
\end{array}
\end{equation*}
It has an  MCB whose periodic orbit bifurcation is a period-doubling at the origin,  a fixed point that becomes a repeller. We find numerically that immediately after the bifurcation  the chaotic attractor has a dense set of repellers and a dense set of saddles. This observation supports Conjecture~3.
For $\alpha=0.07$ and $\sigma\in (0.2, 1/\pi) $, there is a chaotic attractor for which all periodic orbits in the attractor are saddles. Then we increase $\sigma$ so that the origin (which is in the attractor) period-doubles at $\sigma = \sigma_0 = 1/\pi \sim 0.318$ (the MCB value) 
and $(0,0)$ becomes a repeller.
As $\sigma$ increases from beyond $\sigma_0$ a new index-$2$ set appears in the attractor, and repelling periodic orbits are immediately dense in the attractor (Fig.~\ref{fig:Kos-saddles} left), and the saddle periodic orbits are still dense in the attractor (Fig.~\ref{fig:Kos-saddles} right).

{\bf Discussion. } Multi-chaos is important for all models with high-dimensional attractors including weather prediction and climate modeling. 
It is perhaps the unifying concept linking different phenomena observed in numerous numerical simulations of chaotic dynamical systems and physical experiments, such as unstable dimension variability (UDV), on-off intermittency, riddled basins, blowout and bubbling bifurcations. It is also a major cause of shadowing to fail, i.e., for simulated solutions to be non-physical. We have made three conjectures as the beginning of a general theory of multi-chaos.

Multi-chaotic systems are particularly difficult to visualize, so we have introduced some low-dimensional examples as paradigms, 
including one that is perhaps the simplest possible example of multi-chaos (based on the well-known baker map). 

We investigate how multi-chaos arises as a parameter is varied and find that the transition to multi-chaos occurs at a periodic orbit bifurcation.
Because shadowing fails for multi-chaotic systems, detecting the transition from mono-chaos to multi-chaos can be critical for prediction efforts. 

While the UDV condition requires only two orbits of different UD values,
we have focused on the existence of not just these two orbits but much larger index sets which exist in multi-chaotic attractors and make multi-chaos persistent.

Because of the increasing importance of models with high dimensional chaotic attractors, we have tried to create terminology that is easy to use.

{\bf Acknowledgments}. YS was supported by the JSPS KAKENHI Grant No.17K05360 and JST PRESTO JPMJPR16E5.
MAFS was supported by the Spanish State Research Agency (AEI) and the European Regional Development Fund (FEDER)
No.FIS2016-76883-P and jointly by the Fulbright Program and the Spanish Ministry of Education
No.FMECD-ST-2016.

\end{document}